\documentclass[doublecol]{epl2} 
\usepackage{amsmath}

\title{Anomalous Diffusion on the Hanoi Networks}
\author{Stefan Boettcher and Bruno Gon{\c c}alves}
\institute{                    
 Dept. of Physics,  Emory University, Atlanta, GA 30322%\\
}

\abstract{Diffusion is modeled on the recently proposed Hanoi networks
  by studying the mean-square displacement of random walks with time,
  $\left\langle r^{2}\right\rangle \sim t^{2/d_{w}}$.  It is found
  that diffusion -- the quintessential mode of transport throughout
  Nature -- proceeds faster than ordinary, in one case with an exact,
  anomalous exponent $d_{w}=2-\log_2(\phi)=1.30576\ldots$.  It is
  an instance of a physical exponent containing the ``golden ratio''
  $\phi=\left(1+\sqrt{5}\right)/2$ that is intimately related to
  Fibonacci sequences and since Euclid's time has been found to be
  fundamental throughout geometry, architecture, art, and Nature
  itself. It originates from a singular renormalization group fixed
  point with a subtle boundary layer, for whose resolution $\phi$ is
  the main protagonist.  The origin of this rare singularity is easily
  understood in terms of the physics of the process. Yet, the
  connection between network geometry and the emergence of $\phi$ in
  this context remains elusive. These results provide an
    accurate test of recently proposed universal scaling forms for
    first passage times.
}

\begin{document}

\maketitle

\section{Introduction}
\label{introduction}
The study of anomalous diffusion is an integral part in the analysis
of transport processes in complex
materials~\cite{Shlesinger84,Bouchaud90,Metzler04,Bollt05,Condamin07}.
Random environments often slow transport significantly, leading to
sub-diffusive behavior. Much attention has thus been paid to model
sub-diffusion on designed structures with some of the trappings of
disordered materials, exemplified by
Refs.~\cite{Ogielski85,Huberman85,Havlin87,Sibani87,Maritan93,Anh05}. Even
self-organized critical processes can be shown to evolve
sub-diffusively, controlled by the memory of all past
events~\cite{BoPa2}. On the other hand, tracer particles in rapidly
driven fluids may exhibit super-diffusive behavior~\cite{Solomon93},
typically modeled in terms of L\'evy
flights~\cite{Shlesinger93,Metzler04}. Both regimes are self-similar,
fractal generalizations of ordinary diffusion.

In this Letter we consider diffusion on two new networks, which yield
interesting realizations of super-diffusive behavior. Both of these
networks were introduced to explore certain aspects of small-world
behavior~\cite{SWPRL}. Their key distinguishing
  characteristic is their ability to mix a geometric backbone, i.~e. a
  one-dimensional lattice, with small-world links in a non-random,
  hierarchical structure. In particular, these networks permit a
  smooth interpolation between finite-dimensional and mean-field
  properties, which is absent from the renormalization group (RG) due
  to Migdal and Kadanoff, for instance~\cite{Plischke94}. The unusual
  structure of these networks recasts the RG into a novel form, where
  the equations are essentially those of a one-dimensional model in
  which the complex hierarchy enters at each RG-step as a (previously
  unrenormalized) source term. This effect is most apparent in the
  real-space RG for the Ising models discussed in
  Ref.~\cite{SWPRL}. It is obscured in our dynamic RG treatment below,
  since these walks are always embedded on the lattice backbone.  On
  the practical side, their regular, hierarchical structure allows
for easily engineered implementations, say, to efficiently synchronize
communication networks~\cite{SWPRL}. Regarding diffusion, one of the
networks proves to be merely an incarnation of a Weierstrass random
walk found for L\'evy flights~\cite{Shlesinger93} with ballistic
transport, while the other network shows highly non-trivial transport
properties, very much unlike a L\'evy flight, as revealed
by our exact RG treatment. The fixed point equations are singular and
exhibit a boundary layer~\cite{BO}. It provides a tangible case of a
singularity in the RG~\cite{Griffiths78,Maritan93} that is easily
interpreted in terms of the physics.

\section{Generating  Hanoi Networks}
\label{generating}
In the Tower-of-Hanoi problem~\cite{Sedgewick04}, disks of increasing
size, labeled $i=1$ to $k$ from top to bottom, are stacked up and have
to be moved in a Sisyphean task into a 2nd stack, disk-by-disk, while
at no time a larger disk can be placed onto a smaller one. To this
end, a 3rd stack is provided as overflow. First, disk 1 moves to the
overflow and disk 2 onto the 2nd stack, followed by disk 1 on top of
2. Now, disk 3 can move to the overflow, disk 1 back onto disk 4, disk
2 onto 3, and 1 onto 2. Now we have a new stack of disks 1, 2, and 3
in prefect order, and only $k-3$ more disk to go! But note the values
of disk-label $i$ in the sequence of moves:
1-2-1-3-1-2-1-4-1-2-1-3-1-2-1-5-..., and so on.

\begin{figure}
\includegraphics[bb=47bp 50bp 870bp 550bp,clip,scale=0.3]{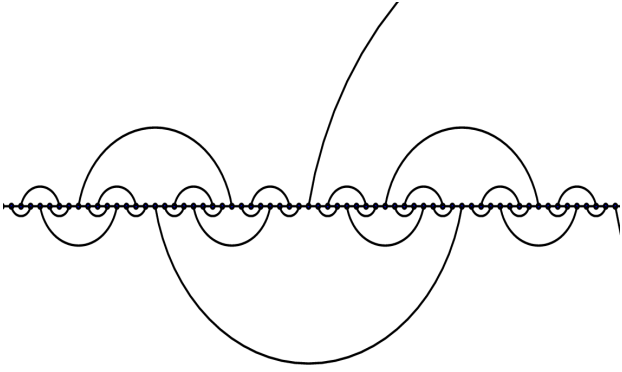}
\caption{Depiction of the planar ``Tower-of-Hanoi'' network HN3. Here, the
  $1d-$backbone of sites extends over $0<n<\infty$.}
\label{fig:3hanoi}
\end{figure}

\begin{figure}
\includegraphics[angle=90,bb=20bp 0bp 450bp 650bp,clip,scale=0.36]{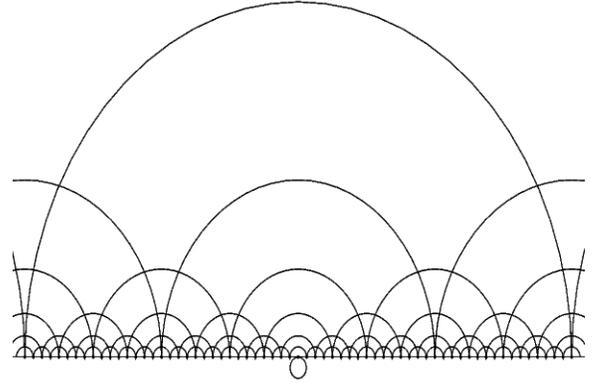}
\caption{Depiction of the {}``Tower-of-Hanoi'' network HN4. Here, the
  $1d-$backbone of sites extends over $-\infty<n<\infty$. The site
  $n=0$, not covered by Eq.~(\ref{neq}), is special and is connected
  to itself here.}
\label{fig:4hanoi}
\end{figure}

Inspired by models of ultra-slow
diffusion~\cite{Ogielski85,Huberman85}, we create our networks as
follows. First, we lay out this sequence on a $1d-$line of
nearest-neighbor connected sites labeled from $n=1$ to $n=L=2^{k}-1$
(the number of moves required to finish the problem). In general, any
site $n(\not=0)$ can be described uniquely by
\begin{equation}
n=2^{i-1}(2j+1),
\label{neq}
\end{equation}
where $i$ is the label of the disk moved at step $n$ in the sequence
above and $j=0,1,2\ldots$. To wit, let us further connect each site
$n$ to the closest site $n'$ that is $2^i$ steps away and possesses
the same value of $i$, both only having a site of value at most $i+1$
between them. According to the sequence, site $n=1$ (with $i=1$) is
now also connected to $n'=3$, 5 to 7, 9 to 11, etc.  For sites with
$i=2$, site $n=2$ now also connects to $n'=6$, 10 to 14, 18 to 22,
etc, and so on also for $i>2$. As a result, we get the network
depicted in Fig.~\ref{fig:3hanoi} that we call HN3. Except at the
boundary, each site now has three neighbors, left and right along the
$1d$ ''backbone'' and a 3rd link to a site $2^i$ steps away. If we
further connect each site also to a fourth site $2^i$ steps in the
other direction and allow $j=0,\pm1,\pm2,\ldots$, we obtain the
network in Fig.~\ref{fig:4hanoi}, called HN4, where each site now has
four neighbors.

These new ``Tower-of-Hanoi'' networks -- a mix of local, geometric
connections and ``small-world''-like long-range jumps -- has
fascinating properties. It is recursively defined with obvious fractal
features. A collection of the structural and dynamic features of HN3 and HN4
are discussed in Ref.~\cite{SWPRL}, such as results for
  Ising models and synchronization.

\section{Diffusion on the Hanoi Networks}
\label{diffusion}
To model diffusion on these networks, we study simple random walks
with nearest-neighbor jumps along the available links, but
  using the one-dimensional lattice backbone as our metric to measure
  distances, which implies a fractal dimension of $d_f=1$. Embedded in
  that space, we want to calculate the non-trivial diffusion exponent
  $d_w$ defined by the asymptotic mean-square displacement 
\begin{equation}
\left\langle  r^{2}\right\rangle \sim t^{2/d_{w}}.
\label{MSDeq}
\end{equation}
A more extensive treatment yielding also first-return probabilities is given
elsewhere~\cite{SWlong}.

First, we consider a random walk on HN4. The
{}``master-equation''~\cite{Redner01} for the probability of the walker
to be at site $n$, as defined in Eq.~(\ref{neq}), at time $t$ is given by
\begin{eqnarray}
{\cal P}_{n,t}&=&
\frac{1-p}{2}\left[{\cal P}_{n-1,t-1}+{\cal P}_{n+1,t-1}\right]\nonumber\\
\nonumber\\
&&\quad+\frac{p}{2}\left[{\cal P}_{n-2^i,t-1}+{\cal P}_{n+2^i,t-1}\right],
\label{eq:4RW}
\end{eqnarray}
where $p$ is the probability to make a long-range jump. (Throughout
this Letter, we considered $p$ uniform, independent of $n$ or $t$). A
detailed treatment of this equation  in terms of generating
  functions is quite involved and proved fruitless, as will be
shown elsewhere~\cite{SWlong}. Instead, we note that the long-time behavior is
dominated by the long-range jumps, as discussed below for HN3. To
simplify matters, we set $p=1/2$ here, although any other finite probability
should lead to the same result. We make an ``annealed''
approximation, i.~e., we assume that we happen to be at some site $n$
in Eq.~(\ref{neq}) with probability $1/2^i$, corresponding to the
relative frequency of such a site, yet independent of update-time or history.
This ignores the fact that in the network geometry a long jump of
length $2^i$ can be followed \emph{only} by another jump of that
length or a jump of unit length, and that many intervening steps are
necessary to make a jump of length $2^{i+1}$, for instance. Here, at
each instant the walker jumps a distance $2^i$ left or right
irrespectively with probability $1/2^i$, and we can write
\begin{eqnarray}
{\cal P}_{n,t} & = &\sum_{n'}T_{n,n'}{\cal P}_{n',t-1}
\label{eq:Transfer}
\end{eqnarray}
with
\begin{eqnarray}
T_{n,n'}&=&\frac{a-1}{2a}\sum_{i=0}^\infty a^{-i}\left(\delta_{n-n',b^i}+\delta_{n-n',-b^i}\right),
\label{eq:T}
\end{eqnarray}
where $a=b=2$. Eqs.~(\ref{eq:Transfer}-\ref{eq:T}) are identical to
the Weierstrass random walk discussed in
Refs.~\cite{Hughes81,Shlesinger93} for arbitrary $1<a<b^2$. There, it
was shown that $d_w=\ln(a)/\ln(b)$, which leads to the conclusion that
$d_w=1$ in Eq.~(\ref{MSDeq}) for HN4, as has been predicted (with
logarithmic corrections) on the basis of numerical simulations in
Ref.~\cite{SWPRL}. These logarithmic corrections are typical for
walks with marginal recurrence, which typically occurs when $d_w=d_f$, such as for ordinary diffusion in two dimensions~\cite{Bollt05}.

For HN3, the master-equation in the bulk reads for
\begin{eqnarray}
{\cal P}_{n,t} & = &
\frac{1-p}{2}\left[{\cal P}_{n-1,t-1}+{\cal P}_{n+1,t-1}\right]+p\,{\cal P}_{n',t-1},\nonumber\\
\nonumber\\
&&\quad n'=\begin{cases}
n+2^{i},&  j~ {\rm even,}\\
\\
n-2^{i},&  j~ {\rm odd,}\end{cases}
\label{eq:3RW}
\end{eqnarray}
with $n$ as in Eq.~(\ref{neq}), and $p$ as before.

\begin{figure}
\includegraphics[bb=0bp 550bp 380bp 750bp,clip,scale=0.6]{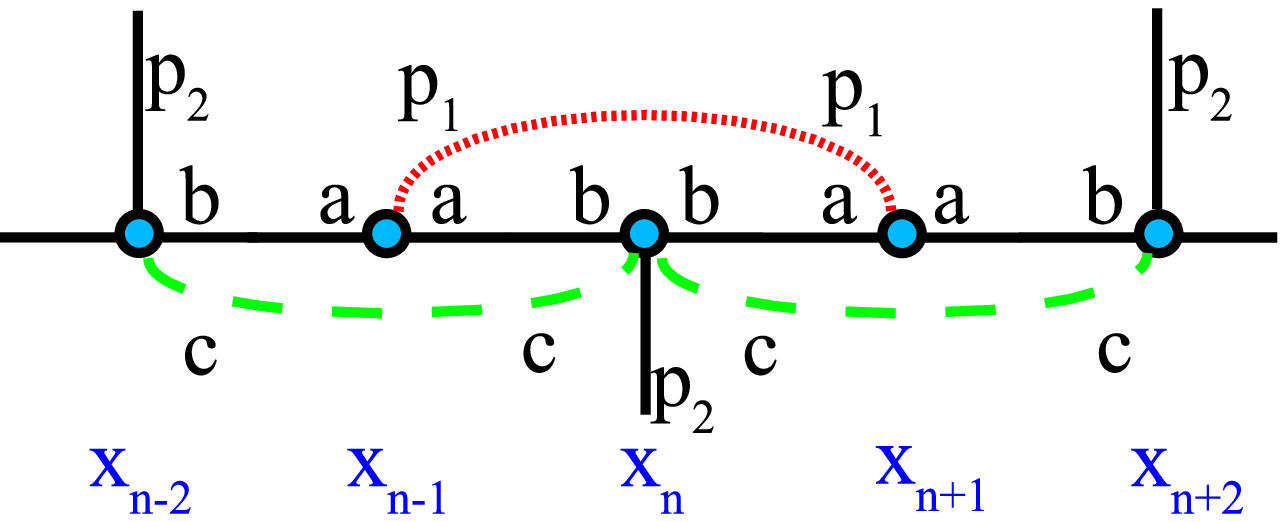}
\includegraphics[bb=0bp 550bp 320bp 750bp,clip,scale=0.6]{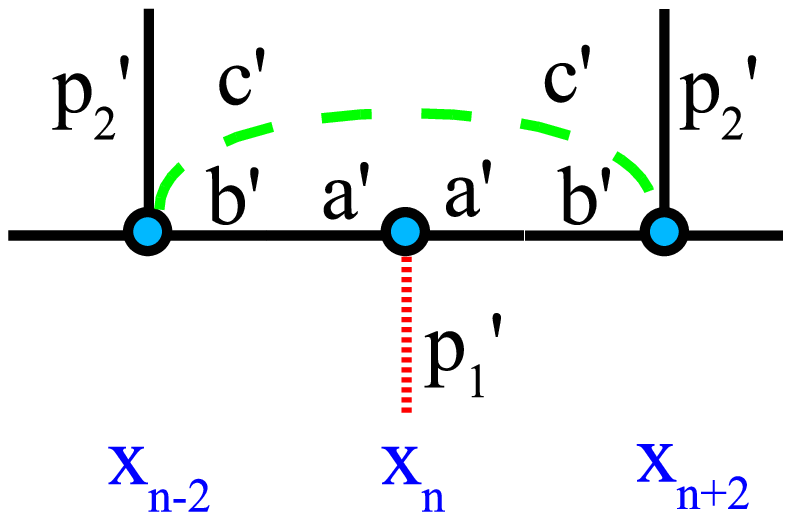}
\caption{Depiction of the (exact) RG step for
random walks on HN3. Hopping rates from one site to another along a
link are labeled at the originating site. The RG step consists of
tracing out odd-labeled variables $x_{n\pm1}$ in the top graph and
expressing the renormalized rates $(a',b',c',p_{1}',p_{2}')$ on the
right in terms of the previous ones $(a,b,c,p_{1},p_{2})$ on the
bottom. The node $x_{n}$, bridged by a (dotted) link between $x_{n-1}$
and $x_{n+1}$, is special as it \emph{must} have $n=2(2j+1)$ and is to
be decimated at the following RG step, justifying the designation of
$p_{1}'$. Note that the original graph in Fig.~\ref{fig:3hanoi} does
not have the green, dashed links with hopping rates $(c,c')$, which
\emph{emerge} during the RG recursion.
}
\label{fig:RG3RW}
\end{figure}

In the RG~\cite{Redner01,Kahng89} solution of Eq.~(\ref{eq:3RW}), at
each step we eliminate all odd sites, i.~e., those sites with $i=0$ in
Eq.~(\ref{neq}). As shown in Fig.~\ref{fig:RG3RW}, the elementary unit
of sites effected is centered at all sites $n$ having $i=1$ in
Eq.~(\ref{neq}). We know that such a site $n$ is surrounded by two
sites of odd index, which are mutually linked. Furthermore, $n$ is
linked by a long-distance jump to a site also of type $i=1$ at $n\pm4$
in the neighboring elementary unit, where the direction does not
matter here. The sites $n\pm2$, which are shared at the boundary
between such neighboring units also have even index, but their value
of $i\geq2$ is indetermined and irrelevant for the immediate RG step, as
they have a long-distance jump to some sites $m_\pm$ at least eight
sites away.

Using a standard generating function~\cite{Redner01},
\begin{eqnarray}
x_{n}(z)&=&\sum_{t=0}^{\infty}{\cal P}_{n,t}\,z^{t},
\label{eq:generator}
\end{eqnarray}
yields for the five
sites inside the elementary unit centered at $n$:
\begin{eqnarray}
x_{n}&=&a\left(x_{n-1}+x_{n+1}\right)\nonumber\\
&&\qquad+c\left(x_{n-2}+x_{n+2}\right)+p_{2}\,x_{n\pm4},
\nonumber\\
\nonumber\\
x_{n\pm1} & = & b\left(x_{n}+x_{n\pm2}\right)+p_{1}\,x_{n\mp1},
\label{eq:transformed3RW}\\
\nonumber\\
x_{n\pm2}&=&a\left(x_{n\pm1}+x_{n\pm3}\right)\nonumber\\
&&\qquad+c\left(x_{n}+x_{n\pm4}\right)+p_{2}\,x_{m_\pm},
\nonumber
\end{eqnarray}
where we have absorbed the parameters $p$ and $z$ into general
{}``hoping rates'' that are initially $a^{(0)}=b^{(0)}=\frac{z}{2}(1-p)$,
$c^{(0)}=0$, and $p_{1}^{(0)}=p_{2}^{(0)}=zp$. 

The RG update step consist of eliminating from these five equations those
two that refer to an odd index, $n\pm1$. After some algebra, we obtain
\begin{eqnarray}
x_{n}&=&b'\left(x_{n-2}+x_{n+2}\right)+p_{1}'\,x_{n\pm4},
\nonumber\\
\nonumber\\
x_{n\pm2}&=&a'\left(x_{n}+x_{n\pm4}\right)\label{eq:RGafter3RW}\\
&&\qquad+c'\left(x_{n\mp2}+x_{n\pm6}\right)+p_{2}'\,x_{m_\pm},
\nonumber
\end{eqnarray}
with
\begin{eqnarray}
a' & = &
\frac{\left[ab+c\left(1-p_{1}\right)\right]\left(1+p_{1}\right)}{1-p_{1}^{2}-2ab},\nonumber\\
\nonumber\\
b' &=&\frac{ab+c\left(1-p_{1}\right)}{1-p_{1}-2ab},\nonumber\\
\nonumber\\
c' &=&\frac{abp_{1}}{1-p_{1}^{2}-2ab},
\label{eq:RG3RWfp}\\  
\nonumber\\
p_{1}'&=&\frac{p_{2}\left(1-p_{1}\right)}{1-p_{1}-2ab},\nonumber\\
\nonumber\\
p_{2}'&=&\frac{p_{2}\left(1-p_{1}^{2}\right)}{1-p_{1}^{2}-2ab}.\nonumber 
\end{eqnarray}
If for all sites $l=n,n\pm2,n\pm4,\ldots$ in Eq.~(\ref{eq:RGafter3RW}) we further identify\footnote{As we will show
elsewhere~\cite{SWlong}, the constant $C$ is determined when initial
and boundary conditions are considered, as is essential for the case of
first transit and return times~\cite{Redner01}.} $x_{l}=C\,x_{l/2}'$,
we note that the primed equations coincide
with the unprimed ones in Eqs.~(\ref{eq:transformed3RW}).  Hence, the
RG recursion equations in (\ref{eq:RG3RWfp}) are \emph{exact} at any
step $k$ of the RG, where unprimed quantities refer to the $k$th
recursion and primed ones to $k+1$.

\begin{figure}
\includegraphics[scale=0.32]{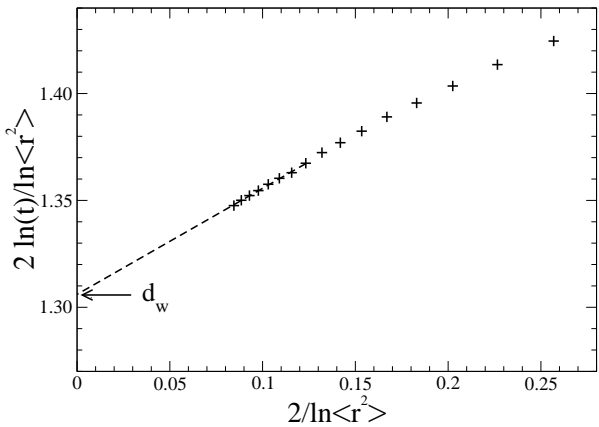}
\caption{Plot of the results from simulations of the mean-square
  displacement of random walks on HN3 displayed in
  Fig.~\ref{fig:3hanoi}. More than $10^7$ walks were evolved up to
  $t_{\rm max}=10^6$ steps to measure $\langle r^2\rangle_t$. The data
  is extrapolated according to Eq.~(\ref{MSDeq}), such that the
  intercept on the vertical axis determines $d_w$ asymptotically. The
  exact result from Eq.~(\ref{eq:D-expo}) is indicated by the arrow.
}
\label{fig:MSDextra}
\end{figure}

\begin{figure}
\includegraphics[scale=0.32]{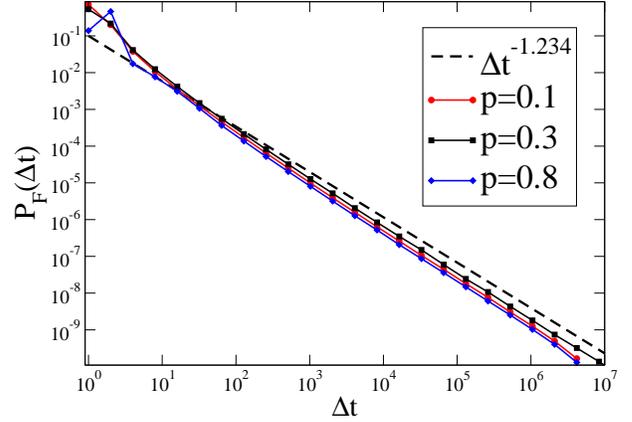}
\caption{Plot of the probability $P_F(\Delta t)$ of first returns to
  the origin after $\Delta t$ update steps on a system of unlimited
  size. Data was collected for three different walks on HN3 with
  $p=0.1$ (circles), $p=0.3$ (squares), and $p=0.8$ (diamonds). The
  data with the smallest and largest $p$ exhibit strong transient
  effects. The exact result in Eq.~(\ref{eq:tau-expo}),
  $\mu=1.234\ldots$, is indicated by the dashed line.
}
\label{fig:FR}
\end{figure}

Solving Eqs.~(\ref{eq:RG3RWfp}) algebraically at infinite time [which
corresponds to the limit $z\nearrow1$, see Eq.~(\ref{eq:generator})]
and for $k+1\sim k\to\infty$ (by dropping the prime on all left-hand
parameters), we -- apparently -- obtain only two fixed points at
$a=b=1/2$ and $c=p_{1}=p_{2}=0$, and $a=b=c=0$ and $p_{1}=p_{2}=1$.
The first fixed point corresponds to an ordinary $1d$ walk without
long-range jumps, in the second there is no hopping along the
$1d$-backbone at all and the walker stays \emph{confined,} jumping
back-and-forth within a single, long-range jump. Yet, both fixed
points are \emph{unstable} with respect to small perturbations in the
initial parameters.

Starting with any positive probability $p$ for long-range jumps, those
dominate over the $1d$ walk at long times. Paradoxically, exclusive
long-range jumps found at the 2nd fixed point lead to confinement,
itself undermined by \emph{any} positive probability to escape along
the $1d-$line, allowing to reach even longer jumps. Instead, the
process gets attracted to a third, stable fixed point hidden inside a
singular \emph{boundary layer}\cite{BO} in the renormalization group
equations~(\ref{eq:RG3RWfp}) near the confined state.

We have to account for the asymptotic boundary layer in
Eqs.~(\ref{eq:RG3RWfp}) with the Ansatz 
$y\sim A_y\alpha^{-k}\to0$ for
$y\in\{a,b,c,1-p_{1},1-p_{2}\}$, where $k\to\infty$
refers to the $k$th RG step.  Choosing $A_a=1$, the other $A_y$'s and
the eigenvalues $\alpha$ are determined
\emph{self-consistently}.
The only eigenvalue
 satisfying the requirement $\alpha>1$  is
$\alpha=2/\phi$. Here, $\phi=\left(\sqrt{5}+1\right)/2=1.6180\ldots$
is the legendary ``golden ratio''~\cite{Livio03} defined by
Euclid~\cite{Euclid}. Hence, every renormalization of network size,
$L\to L'=2L$, has to be matched by a rescaling of hopping rates with
$\alpha=2/\phi$ to keep motion along the $1d$-backbone finite and
prevent confinement.

Extending the analysis to include finite-time corrections
(i.~e., $1-z\ll1$), we extend the above Ansatz to
\begin{eqnarray} 
y^{(k)}&\sim& A_{y}\alpha^{-k}\left\{1+\left(1-z\right)B_{y}\beta^{k}+\ldots\right\}
\label{eq:Ansatz}
\end{eqnarray}
for all $y\in\{a,b,c,1-p_{1},1-p_{2}\}$. In addition to the
leading-order constants $A_{y}$ and $\alpha$, also the next-leading
constants are determined self-consistently, and we extract uniquely
$\beta=2\alpha$. Accordingly, time re-scales now as
\begin{eqnarray} 
T &\to & T'=2\alpha T,
\label{eq:Tscal}
\end{eqnarray}
and we obtain from Eq.~(\ref{MSDeq}) with $T\sim L^{d_{w}}$
for the diffusion exponent for HN3
\begin{eqnarray} 
d_{w}&=&2-\log_2\phi=1.30576\ldots.
\label{eq:D-expo}
\end{eqnarray}
The result for $d_{w}$ is in excellent agreement with our simulations,
as shown in Fig.~\ref{fig:MSDextra}.

Using the methods from Ref.~\cite{Redner01}, a far more
  extensive treatment shows~\cite{SWlong} that the exponent $\mu$
for the probability distribution, $P_F(\Delta t)\sim\Delta
t^{-\mu}$, of first-return times $\Delta t$ is given by
\begin{eqnarray} 
\mu&=&2-\frac{1}{d_w}=1.2342\ldots.
\label{eq:tau-expo}
\end{eqnarray}
The relation between $\mu$ and $d_w$ is typical also for L\'evy
flights~\cite{Metzler04}, and the result is again borne out by our
simulations, see Fig.~\ref{fig:FR}. It is remarkable,
  though, that the more detailed analysis in Ref.~\cite{SWlong} also
  shows that walks on HN3 are \emph{not} uniformly recurrent, as the
  result of $d_w>d_f=1$ here would indicate. That calculation shows
  that only sites on the highest level of the hierarchy are
  recurrent. While all other sites do share the same exponent $\mu$
  in Eq.~(\ref{eq:tau-expo}) for actual recurrences, they have a
  diminishing return probability with decreasing levels in the
  infinite system limit. This is clearly a consequence of walkers
  being nearly-confined to the highest levels of the hierarchy at long
  times, as expressed by the boundary layer.

We finally contrast the behavior of HN4 discovered above with the
analysis of HN3. Clearly, when long-range jumps are interconnected as
in HN4, there is no confinement, the boundary layer disappears [which
  would be similar to $\alpha=1$ in
  Eqs.~(\ref{eq:Ansatz}-\ref{eq:Tscal}) for HN3], and diffusion
spreads ballistically, $d_{w}=1$. Our numerical studies, and the
similarity to Weierstrass random walks~\cite{Hughes81}, further
supports that $\mu$ for walks on HN4 is also given by
Eq.~(\ref{eq:tau-expo}), leading to $\mu=1$. This scaling is
  again indicative of a marginally recurrent state and requires
  logarithmic corrections for proper normalization, as was observed in
  simulations~\cite{SWPRL}. 

\section{Conclusions}
\label{conclusions}
We conclude with two further considerations. First, in reference to
the potential of these networks to interpolate between long-range and
a finite-dimensional behavior that we invoked in the introduction, we
just add the following illustrative remark: If the probability to
undertake a long-distance jump would be distance-dependent in each
level of the hierarchy, we can obtain immediately a new result for
walks on HN4 in the annealed approximation above. Let $p$ vary with a
power of the backbone-distance between sites, say $p\propto
r^{-\sigma}$, then for each level $i$ of the hierarchy it is
$r=r_i=2^i$, i.~e. $p=p_i\propto2^{-i\sigma}$, and the weight to make
a jump of length $2^i$ in Eq.~(\ref{eq:T}) is given by
$a=2^{1+\sigma}$, leading to $d_w=1+\sigma$. As can be expected, the
analysis of the Weierstrass walk breaks down for $\sigma\to1^-$, at
which point the long-range jumps become irrelevant and we obtain the
results for ordinary $1d$ diffusion. Hence, $0\leq\sigma\leq1$
interpolates analytically between long-range and one-dimensional
behavior of the random walk on HN4. (In fact, the analysis formally
can be extended to $0>\sigma>-1$, where the walk becomes non-recurrent
and is dominated by high levels in the hierarchy. Yet, the annealed
approximation that assumes free transitions between different levels
of the hierarchy is bound to fail.)

\begin{figure}
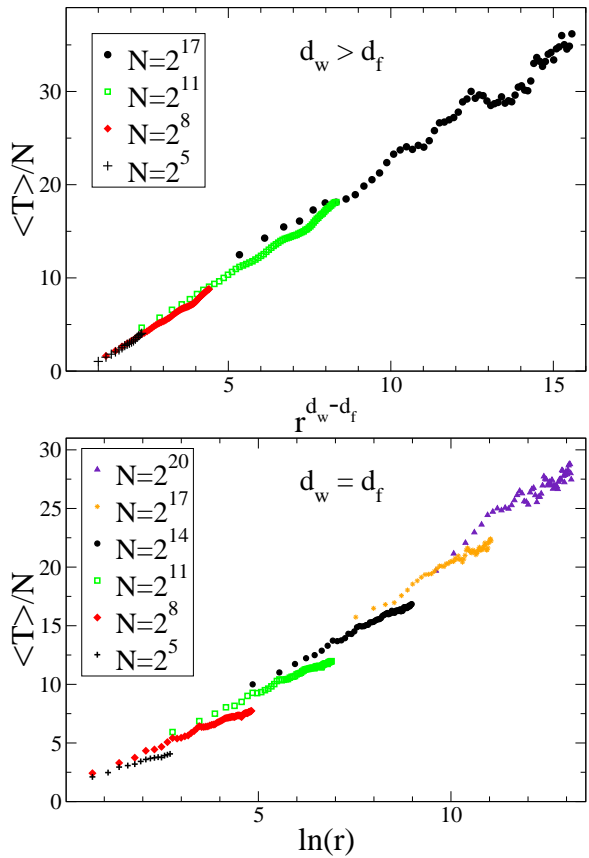

\includegraphics[scale=0.32]{mfpt_HN3linear.eps}

\includegraphics[scale=0.32]{mfpt_HN4center.eps}
\caption{Plot of the mean first-passage times $\langle T\rangle$ for
  walks as a function of distance $r$ between starting and target site
  for HN3 (top) and HN4 (bottom). The data has been scaled according
  to Eq.~(\ref{eq:condamin} such that the data collapses
  asymptotically onto a line that only depends on the model but that
  is independent of system size $N$. This collapse is excellent for
  HN3,  it is somewhat weaker for HN4. Although all system sizes lead
  to linear forms in $\ln(r)$, their slope apparently varies with $N$.
  This could be caused by logarithmic scaling corrections to the
  slope, or by the lack of asymptotic behavior at the available system
  sizes $N$.  }
\label{fig:mfpt}
\end{figure}

Our final consideration concerns a recent proposal by Condamin et
al~\cite{Condamin07} for a very general scaling form for mean
first-passage times $\langle T\rangle$ for walks as a function of
distance $r$ between starting and target site on a graph (lattice,
network, etc.) of $N$ sites. Based on $d_w-d_f$,
Ref.~\cite{Condamin07} determined that
\begin{eqnarray}
\langle T\rangle\sim N\begin{cases} A + B r^{d_w-d_f},& d_w>d_f,\\ A +
B \ln(r),& d_w=d_f,\\ A-Br^{d_w-d_f},&d_w<d_f,
\end{cases}
\label{eq:condamin}
\end{eqnarray}
for \emph{fixed} constants $A,B$, independent of $N$ and $r$. Our
networks provide a non-trivial set of exponents to explore these
relations with simple simulations. In particular, HN3 with
$d_w-d_f=0.30576$ provides an instance for a powerlaw-divergent mean
first-passage time, while HN4 exactly probes the marginal case
$d_w=d_f(=1)$ with a logarithmic divergence of $\langle
T\rangle$. When plotting $\langle T\rangle/N$ as a function of
$r^{d_w-d_f}$ or $\ln(r)$, resp., in Figs.~\ref{fig:mfpt} we indeed
obtain a universal straight line over many orders of magnitude in $N$
and $r$, indicative of fixed $A,B$.

\acknowledgments We like to thank F.~Family, S.~Redner,
S.~Coppersmith, and M.~Shlesinger for helpful discussions. We thank
the referee for calling our attention to Ref.~\cite{Condamin07}.

\bibliographystyle{unsrt}
\bibliography{/Users/stb/Boettcher}

\begin{thebibliography}{10}

\bibitem{Shlesinger84}
M.~F. Shlesinger and B.~J. West, editors.
\newblock {\em Random walks and their applications in the physical and
  biological sciences}.
\newblock American Institute of Physics, New York, 1984.

\bibitem{Bouchaud90}
Jean-Philippe Bouchaud and Antoine Georges.
\newblock Anomalous diffusion in disordered media: Statistical mechanisms,
  models and physical applications.
\newblock {\em Physics Reports}, 195:127--293, 1990.

\bibitem{Metzler04}
R.~Metzler and J.~Klafter.
\newblock The restaurant at the end of the random walk: recent developments in
  the description of anomalous transport by fractional dynamics.
\newblock {\em J. Phys. A: Math. Gen.}, 37:R161--R208, 2004.

\bibitem{Bollt05}
E.~M. Bollt and D.~ben Avraham.
\newblock What is special about diffusion on scale-free nets?
\newblock {\em New Journal of Physics}, 7:26, 2005.

\bibitem{Condamin07}
S.~Condamin, O.~Benichou, V.~Tejedor, R.~Voituriez, and J.~Klafter.
\newblock First-passage times in complex scale-invariant media.
\newblock {\em Nature}, 450:77, 2007.

\bibitem{Ogielski85}
Andrew~T. Ogielski and D.~L. Stein.
\newblock Dynamics on ultrametric spaces.
\newblock {\em Phys. Rev. Lett.}, 55(15):1634--1637, Oct 1985.

\bibitem{Huberman85}
B~A Huberman and M~Kerszberg.
\newblock Ultradiffusion: the relaxation of hierarchical systems.
\newblock {\em J. Phys. A: Math. Gen.}, 18(6):L331--L336, 1985.

\bibitem{Havlin87}
S.~Havlin and D.~Ben-Avraham.
\newblock Diffusion in disordered media.
\newblock {\em Adv. Phys.}, 36:695--798, 1987.

\bibitem{Sibani87}
P.~Sibani and K.-H. Hoffmann.
\newblock Random walks on cayley trees: Temperature-induced
  transience-recurrence transition, small exponents and logarithmic relaxation.
\newblock {\em Europhys. Lett.}, 4:967--972, 1987.

\bibitem{Maritan93}
A.~Maritan, G.~Sartoni, and A.~L. Stella.
\newblock Singular dynamical renormalization group and biased diffusion on
  fractals.
\newblock {\em Phys. Rev. Lett.}, 71(7):1027--1030, Aug 1993.

\bibitem{Anh05}
Do~Hoang~Ngoc Anh, K.~H. Hoffmann, S.~Seeger, and S.~Tarafdar.
\newblock Diffusion in disordered fractals.
\newblock {\em Europhys. Lett.}, 70:109--115, 2005.

\bibitem{BoPa2}
S.~Boettcher and M.~Paczuski.
\newblock Ultrametricity and memory in a solvable model of self-organized
  criticality.
\newblock {\em Phys. Rev. E}, 54:1082, 1996.

\bibitem{Solomon93}
T.~H. Solomon, E.~R. Weeks, and H.~L. Swinney.
\newblock Observation of anomalous diffusion and lévy flights in a
  two-dimensional rotating flow.
\newblock {\em Phys. Rev. Lett.}, 71:3975 -- 3978, 1993.

\bibitem{Shlesinger93}
M.~F. Shlesinger, G.~M. Zaslavsky, and J.~Klafter.
\newblock Strange kinetics.
\newblock {\em Natur}, 363:31--37, 1993.

\bibitem{SWPRL}
S.~Boettcher, B.~Gon{\c c}alves, and H.~Guclu.
\newblock Hierarchical regular small-world networks.
\newblock {\em J. Phys. A: Math. Theor.}, 41:252001, 2008.
\newblock arxiv:0712.1259.

\bibitem{Plischke94}
M.~Plischke and B.~Bergersen.
\newblock {\em Equilibrium Statistical Physics, 2nd edition}.
\newblock World Scientifc, Singapore, 1994.

\bibitem{BO}
C.~M. Bender and S.~A. Orszag.
\newblock {\em Advanced Mathematical Methods for Scientists and Engineers}.
\newblock McGraw-Hill, New York, 1978.

\bibitem{Griffiths78}
R.~B. Griffiths and P.~A. Pearce.
\newblock Position-space renormalization-group transformations: Some proofs and
  some problems.
\newblock {\em Phys. Rev. Lett.}, 41:917--920, 1978.

\bibitem{Sedgewick04}
R.~Sedgewick.
\newblock {\em Algorithms in C, 3rd Edition}.
\newblock Addison-Wesley, Boston, 2004.

\bibitem{SWlong}
Stefan Boettcher, Bruno Goncalves, and Julian Azaret.
\newblock Geometry and dynamics for hierarchical regular networks.
\newblock {\em Journal of Physics A: Mathematical and Theoretical},
  41(33):335003, 2008.

\bibitem{Redner01}
S.~Redner.
\newblock {\em A Guide to First-Passage Processes}.
\newblock Cambridge University Press, Cambridge, 2001.

\bibitem{Hughes81}
B.~D. Hughes, M.~F. Shlesinger, and E.~W. Montroll.
\newblock Random walks with self-similar clusters.
\newblock {\em Proc. Natl. Acad. Sci.}, 78:3287--3291, 1981.

\bibitem{Kahng89}
B~Kahng and S~Redner.
\newblock Scaling of the first-passage time and the survival probability on
  exact and quasi-exact self-similar structures.
\newblock {\em J. Phys. A: Math. Gen.}, 22:887--902, 1989.

\bibitem{Livio03}
M.~Livio.
\newblock {\em The Golden Ratio: The Story of PHI, the World's Most Astonishing
  Number}.
\newblock Broadway Books, New York, 2003.

\bibitem{Euclid}
Euclid.
\newblock Elements, {B}ook {VI}, {D}efinition 3.
\newblock c. 300BC.

\end{thebibliography}

\end{document}